\pdfoutput=1
\documentclass[runningheads]{llncs}
\usepackage{graphicx}
\usepackage[boxed, linesnumbered]{algorithm2e}
\usepackage{amssymb}
\usepackage{amsmath}
\usepackage{dsfont}
\usepackage{url}
\usepackage{subcaption}
\usepackage{cite}

\newcommand\blfootnote[1]{%
  \begingroup
  \renewcommand\thefootnote{}\footnote{#1}%
  \addtocounter{footnote}{-1}%
  \endgroup
}

\begin{document}
\title{A Combination of Theta*, ORCA and Push and Rotate for Multi-agent Navigation
%\thanks{Supported by organization x.}
}

\titlerunning{A Combination of Theta*, ORCA and Push and Rotate ...}

\author{Stepan Dergachev\inst{1, 2}\orcidID{0000-0001-8858-2831} \and \\
Konstantin Yakovlev\inst{2, 1}\orcidID{0000-0002-4377-321X} \and \\
Ryhor Prakapovich\inst{3}\orcidID{0000-0002-3412-9174}}

% \orcidID{1111-2222-3333-4444}

\authorrunning{S. Dergachev et al.}

\institute
{
    National Research University Higher School of Economics, Moscow, Russia
    \and
    Federal Research Center for Computer Science and Control of Russian Academy of Sciences, Moscow, Russia
    \and
    United Institute of Informatics Problems of the National Academy of Sciences of Belarus, Minsk, Belarus\\
    \email{sadergachev@edu.hse.ru, yakovlev@isa.ru, rprakapovich@robotics.by}
}

\maketitle              % typeset the header of the contribution
\begin{abstract}
We study the problem of multi-agent navigation in static environments when no centralized controller is present. Each agent is controlled individually and relies on three algorithmic components to achieve its goal while avoiding collisions with the other agents and the obstacles: \emph{i}) individual path planning which is done by \textsc{Theta*} algorithm; \emph{ii}) collision avoidance while path following which is performed by \textsc{ORCA*} algorithm; \emph{iii}) locally-confined multi-agent path planning done by \textsc{Push and Rotate} algorithm. The latter component is crucial to avoid deadlocks in confined areas, such as narrow passages or doors. We describe how the suggested components interact and form a coherent navigation pipeline.
We carry out an extensive empirical evaluation of this pipeline in simulation. The obtained results clearly demonstrate that the number of occurring deadlocks significantly decreases enabling more agents to reach their goals compared to techniques that rely on collision-avoidance only and do not include multi-agent path planning component \blfootnote{This is a preprint of the paper accepted to ICR'20}.

\end{abstract}
\section{Introduction}

When a group of mobile agents, such as service robots or video game characters, operates in the shared environment one of the key challenges they face is arriving at their target locations while avoiding collisions with the obstacles and each other. In general two approaches to solve this problem, i.e. multi-agent navigation, can be identified: centralized and decentralized.

The centralized approaches rely on a central planner that constructs a joint multi-agent plan utilizing the full knowledge of the agents' states. When the plan is constructed each agent follows it and if the execution is perfect or near-perfect collisions are guaranteed not to happen. Centralized planning can be carried out in such a way that the joint plan is guaranteed to be optimal\footnote{Typically, optimal centralized planers rely on the discretization of the workspace thus the solutions they provide are optimal w.r.t. the used space/time discretization.}. Non-optimal yet computationally efficient centralized planners do exist as well.

\begin{figure}[t]
    \centering
    \includegraphics[width=0.55\linewidth]{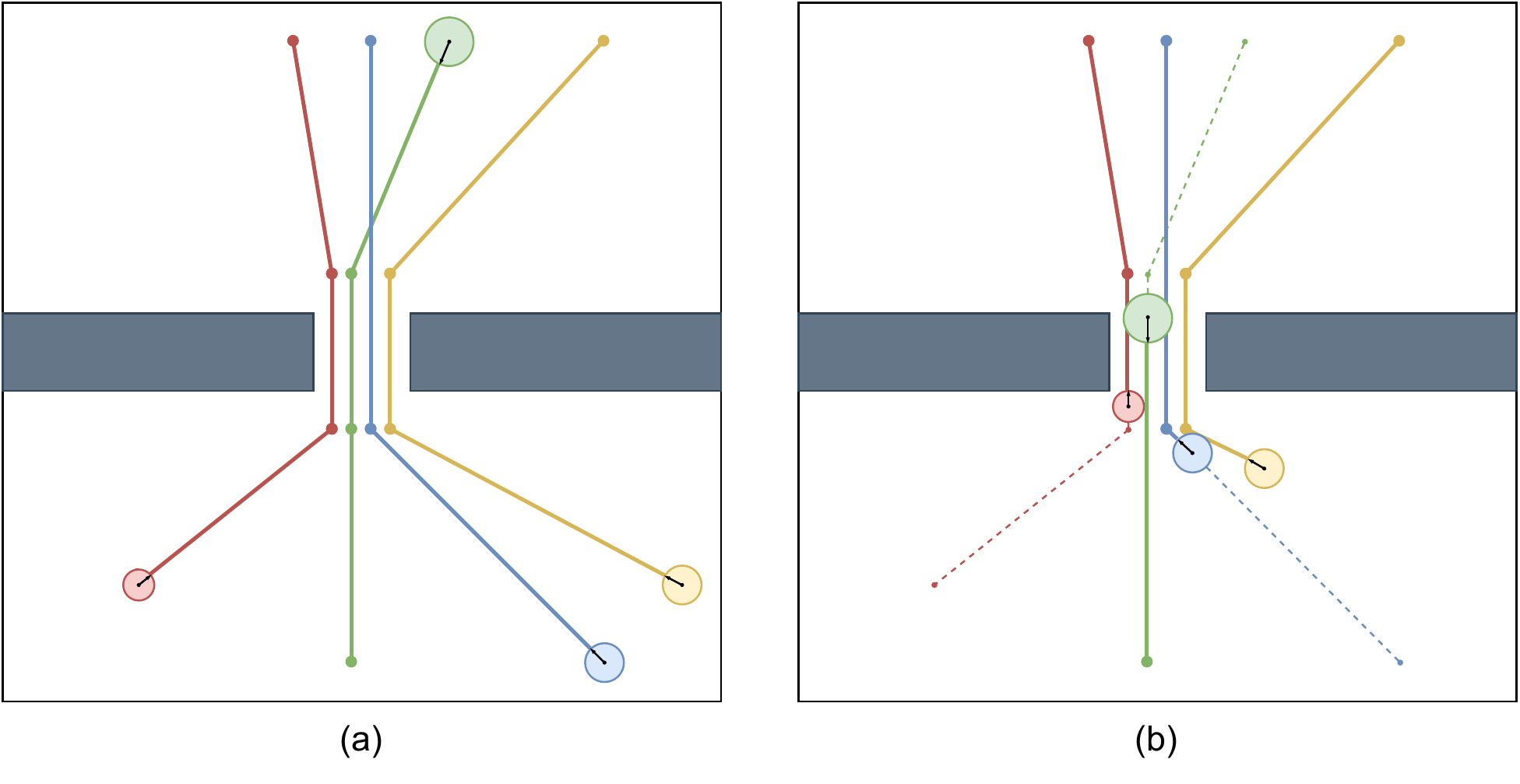}
    \caption{An example of possible deadlock in the vicinity of narrow passage.} 
    \label{fig:intro}
\end{figure}

Decentralized approaches do not rely on the central planner. Instead each agent plans its trajectory individually and resolves conflicts at the execution stage, i.e. an agent reactively avoids collisions relying on the local observations and/or communication. Such approach naturally fits to numerous scenarios with heterogeneous agents, limited communication/visibility etc. It also scales well to large number of agents however it does not guarantee optimality. Moreover, no guarantees that each agent will reach its goal can be provided. The main reason for that is that each agent is egoistic in pursuing its goal this can lead to a deadlock. Consider a scenario shown on \figurename~\ref{fig:intro} when agents have to swap sides using a tight passage connecting the rooms. If they approach the passage nearly at the same time they will be stuck. For all of them to safely move through at least one of the agents has to step back and let other agent go. Thus, some form of cooperative behaviour (which is explicitly or implicitly centralized) is vital for accomplishing the mission.

In this work we present an algorithmic pipeline that combines decentralized and centralized approaches to multi-agent navigation. Within this pipeline each agent constructs and follows its path individually. When a pattern leading to potential deadlock is detected a locally-confined multi-agent path finding is invoked, i.e. agents that appear to be involved in a deadlock are identified and a joint safe plan for these agents is constructed. After executing this plan all involved agents switch back to the decentralized mode. We use \textsc{Theta*}~\cite{daniel2010theta}, \textsc{ORCA*}~\cite{van2011reciprocal}, \textsc{Push and Rotate}~\cite{de2014push} for individual path planing, local collision avoidance and multi-agent path finding respectively. Meanwhile, the suggested pipeline is not limited to these algorithms and other algorithmic components may be used instead of the aforementioned methods as well. We extensively study the suggested approach in simulation and show that the suggested approach leads to much better results compared to fully decentralized method that does not utilize multi-agent path finding.

\section{Related Work}

The approaches to multi-agent navigation can be roughly divided into the two groups: centralized and decentralized \cite{xuan2002multi}. The centralized approaches assume that a central planner is available that possess all the information about the agents and the environment and plans for a joint collision-free solution. Naive application of such search algorithms as \textsc{A*}~\cite{hart1968formal}, i.e. planning in the full joint configuration space, which is a cardinal product of the individual configuration spaces, is impractical as the search space grows exponentially in the number of agents. However there exist techniques that greatly reduce the search effort, such as subdimensional expansion~\cite{wagner2015subdimensional}, conflict-based search~\cite{sharon2015conflict} to name a few. The latter technique is very popular nowadays and there exist a large variety of CBS planners that improve the performance of the original algorithm or extend it in numerous ways, e.g. in~\cite{barer2014suboptimal} a sub-optimal version of \textsc{CBS} is presented, the work~\cite{andreychuk2019multi} planning in continuous time is considered etc. All these algorithms are designed with the aim of minimizing cost of the result solution. At the same time, when large cost of the solution is acceptable one can use extremely fast polynomial algorithms, e.g. \textsc{Push and Rotate}~\cite{de2014push}. Finally, a prioritized approach~\cite{vcap2015prioritized, yakovlev2019prioritized} is also a common choice in numerous cases. Prioritized planners are fast and typically provide solutions of reasonably low cost (very close to optimal in many cases). Their main drawback is that they are incomplete in general.

The decentralized approaches do not rely on the full knowledge of the agents' states and the environment and assume that each agent utilizes only the information available to it locally~\cite{dimarogonas2007decentralized, xuan2002multi}. One of the common ways to solve multi-agent navigation problem under a decentralized setting is to plan individual paths for each agent, e.g. by a heuristic search algorithm such as \textsc{Dijkstra}~\cite{dijkstra1959note}, \textsc{A*}~\cite{hart1968formal}, \textsc{Theta*}~\cite{daniel2010theta}, and then let all agent follow their paths. To avoid collisions algorithms of the ORCA-family are typically used~\cite{van2011reciprocal, snape2014smooth, snape2010smooth, alonso2013optimal}. Another techniques that rely on buffered Voronoi cells~\cite{zhou2017fast} or various reinforcement learning techniques~\cite{chen2017decentralized, long2018towards} to avoid collisions can also be employed. In general decentralized approaches are prone to deadlocks and do not guarantee that each agent will reach its destination.

In this work we combine both decentralized and centralized methods into a coherent multi-agent navigation pipeline striving to combine the best of two worlds: flexibility and scalability of the decentralized navigation and completeness of the centralized planning.

\section{Problem Statement}

Consider a set of $n$ agents moving through 2D workspace, $W \in \mathds{R}^2$, composed of the free space and the obstacles: $W=W_{free} \cup O$. Each obstacle, $o \in O$, is a bounded subset in $W$. Every agent, $i$, is modelled as a disk of radius $r_i$. The state of an agent is defined by its position and velocity: $(\mathbf{p}_i, \mathbf{V}_i)$. The latter is bounded:  $||\mathbf{V}_i|| \leq V_i^{max} \in \mathds{R}^+$, i.e. maximum speed of each agent is given.

Let $T=0, \Delta t, 2 \Delta t, ...$ be the discrete time ($\Delta t = const$). For the sake of simplicity assume that $\Delta t = 1$, thus the timeline is $T=0, 1, 2, ...$. At each moment of time an agent can either move or wait (the latter can be considered as moving with the velocity being zero). We neglect inertial effects and assume that an agent moves from one location to the other following a straight line defined by its current velocity. We also assume perfect localization and execution.

At each time step agent knows \emph{i}) its own position, \emph{ii}) positions of all the obstacles, \emph{iii}) states of the neighboring agents. The latter are the agents that are located withing a radius $R_i$. This radius models a communication range of an agent, i.e. we assume that an agent $i$ can communicate with other neighboring agents and acquire any information from them. For the sake of exposition we will refer to this range as to the visibility range further on. 

A spatio-temporal path for an agent is, formally, a mapping: $\pi: T \rightarrow W_{free}$. It can also be represented as a sequence of agent's locations at each time step: $\pi=\{\pi_0, \pi_1, ...\}$. In this work we are interested in converging paths, i.e. paths by which an agent reaches a particular location and never moves away from it. This can be formulated as follows: $\exists k \in T: \pi_{k+t} = \pi_k \: \forall t>1$. 
%Time moment, $k$, by which an agent reaches its final destination following a path defines the cost of this path: $c(\pi)=k$.

The path for the agent $i$ is valid w.r.t. velocity constraints and static obstacles \emph{iff} the following conditions hold:
\begin{equation*}
  \begin{split}
        ||\pi_{t+1} - \pi_{t} || \leq V_i^{max} \Delta t, \; \forall t \in \overline{0,k-1}
        \\
        \rho (\pi_t, o) > r_i, \; \forall o \in O, \; \forall t \in \overline{0,k},
    \end{split}
\end{equation*}

where $\rho (\pi_t, o)$ stands for the distanceto the closest obstacle.

Consider now the two paths for distinct agents: $\pi^i, \pi^j$. They are called conflict-free if they are valid and the agents following them never collide, that is: 
\begin{equation*}
        ||\pi^i_{t} - \pi^j_{t'} || \leq r_i + r_j, \; \forall t \in \overline{0,k^i}, t' \in \overline{0,k^j}.
\end{equation*}

The set of paths $\Pi = \{\pi^1, ..., \pi^n\}$, one for each agent, is conflict-free if any pair of paths forming this set is conflict-free.

\textbf{The problem} we are considering in this paper can now be formulated as follows. Given $n$ start and goal locations at each time step for each agent choose a velocity s.t. the agent reaches the goal at some time step, the agent never collides with static obstacles and other agents, that is the resultant set of paths $\Pi$ is conflict-free. %An illustration of the problem is presented on~\figurename~\ref{fig:problem}

% To evaluate the quality of a solution, the following metrics are considered. \textit{Makespan} -- the time by which the last agent reaches its goal (maximum over the individual costs); \textit{Flowtime} -- the sum of the individual costs.
% \[ Makespan = \underset{i }{\operatorname{max}} \{k^i\},\]
% \[Flowtime = \sum_{i} k^i.\]

% In this work we are not aimed at getting optimal solutions, however, lower-cost solutions are preferable.

\section{Method}
Our approach relies on three algorithmic components: individual planning, path following with collision avoidance and locally-confined multi-agent path finding in case of potential deadlocks. We will now describe all these components.

\subsection{Individual Path Finding}

\begin{figure}[t]
    \centering
    \includegraphics[width=0.55\textwidth]{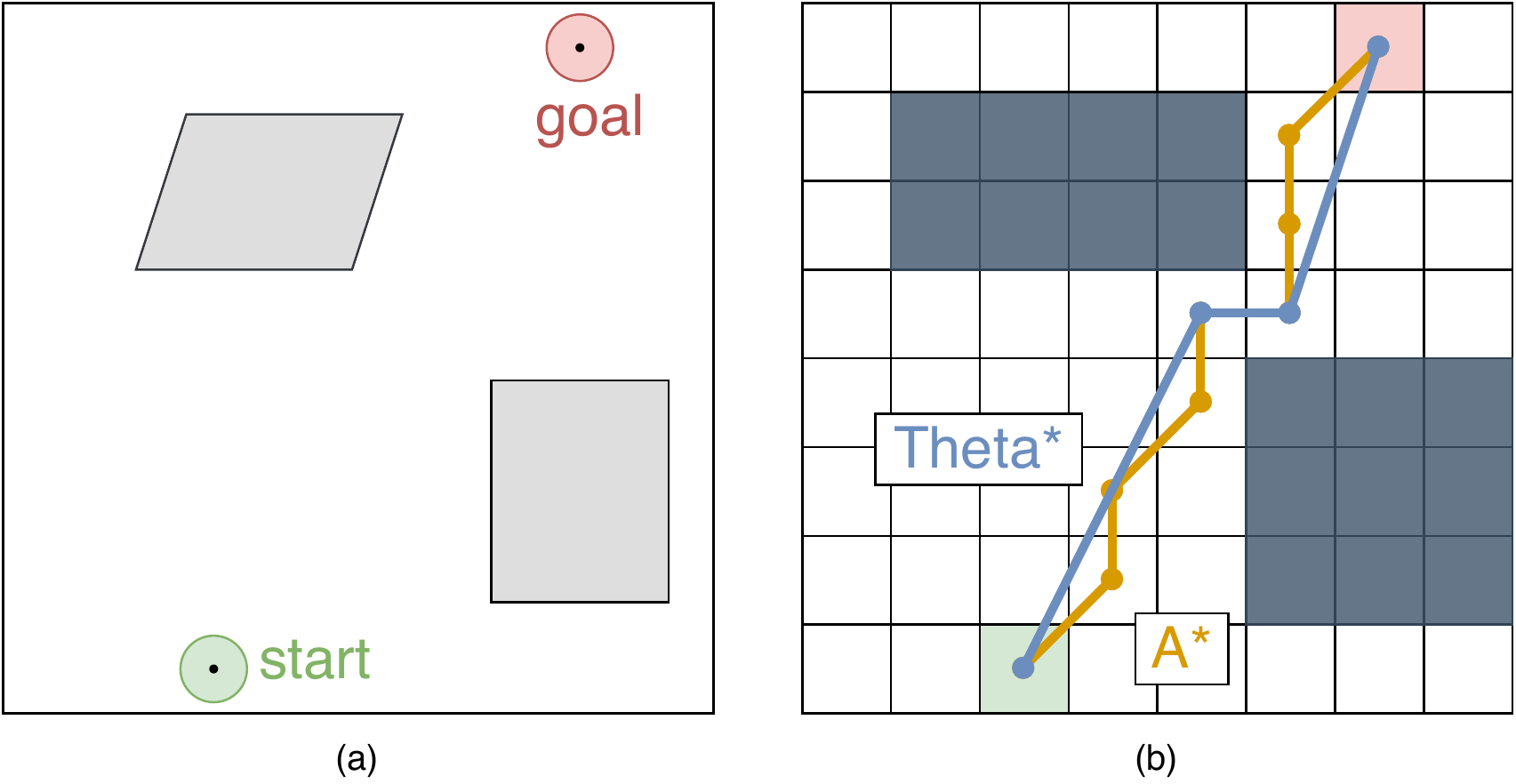}
    \caption{Finding a path for an individual agent: a) the workspace of an agent; b) tesselation of the workspace into the grid and two paths on that grid: the one that was found by \textsc{A*} (in orange) and the one that was found by \textsc{Theta*} (in blue). We use \textsc{Theta*} paths in this work.} 
    \label{fig:traj}
\end{figure}

To find a valid path for an agent that avoids static obstacles we reduce path planning problem to graph search. We employ regular square grids~\cite{yap2002grid} as the discretized workspace representation as they are easy-to-construct and informative graph models widely used for navigation purposes. To create a grid we tessellate the workspace into the square cells and mark each cell either blocked or traversable depending on whether this cell overlaps with any of the obstacle or not (see~\figurename~\ref{fig:traj} on the right). To find a path on the grid we utilize \textsc{Theta*}~\cite{daniel2010theta} algorithm. It's a heuristic search algorithm of the \textsc{A*} family which augments the edge set of the graph on-the-fly by allowing any-angle moves, i.e. the moves between non-adjacent vertices, conditioned that such moves do not lead to the collisions with the obstacles. As a result, paths found by \textsc{Theta*} are typically shorter than the ones found by \textsc{A*} and contain less waypoints. The difference between these two types of paths is clearly visible on~\figurename~\ref{fig:traj}. Both \textsc{Theta*} and \textsc{A*} paths are collision-free, while the former contains 7 intermediate waypoints and the latter -- only 2 of them.

\subsection{Path Following with Collision Avoidance.}

Having the individual paths for all agent the considered multi-agent navigation problem can be naively solved as follows. At each time step an agent chooses such a velocity that it is directed towards the next waypoint on a path and its magnitude is maximal. In other words, agents move along the constructed paths as fast as they can without any temporal waits and spatial deviations. Obviously, such an approach leads to numerous collisions in practice. Instead, we rely on a more advanced method called Optimal Reciprocal Collision Avoidance (\textsc{ORCA})~\cite{van2011reciprocal}. This method is designed to move an agent safely towards its local goal (current waypoint of a path) by reactively adjusting the velocity profile at each timestep. When choosing a velocity \textsc{ORCA} relies on local observations: only other moving agents that are located within a visibility range and static obstacles are taken into account. We tie together individual path planning and path following with collision avoidance in the following fashion.

\begin{figure}[t]
    \centering
    \includegraphics[width=\textwidth]{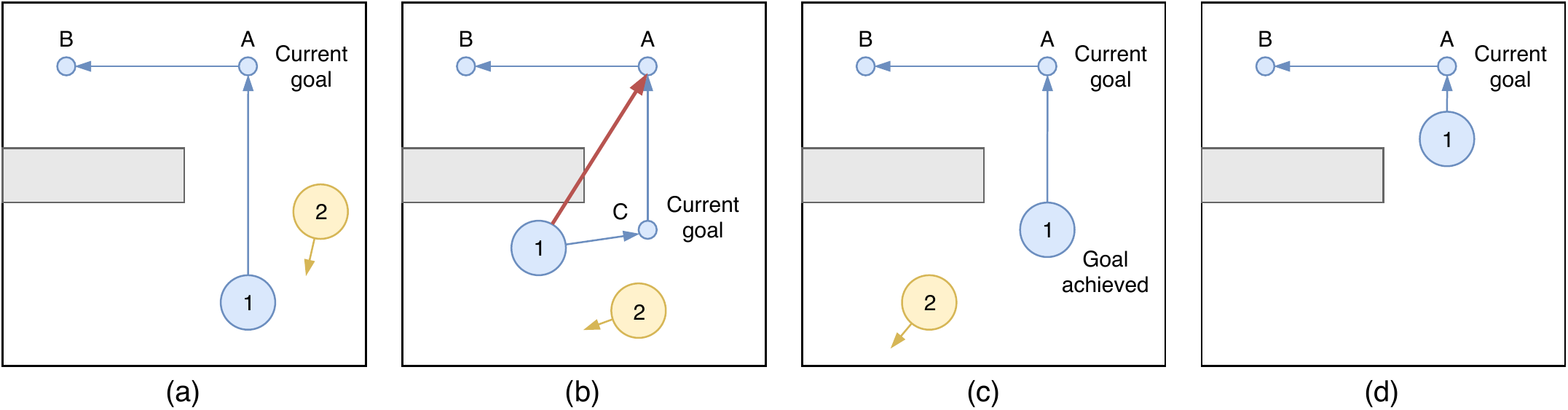}
    \caption{Goal-setting for \textsc{ORCA}. (a) Waypoint $A$ is the local goal of the agent 1. (b) The agent 1 deviates from the original trajectory in order to avoid collision with the agent 2, the line-of-sight to the local goal is lost, re-planning is triggered and the new waypoint, $C$, is added to the path, it becomes the current local goal. (c) $C$ is reached, next local goal is $A$. (d) The agent safely moves towards $A$ without any deviation as no other moving agents interfere with its plan.} 
    \label{fig:currgoal}
\end{figure}

The waypoints that comprise an individual path (the one found by \textsc{Theta*}) form a sequence of the local goals for \textsc{ORCA} and an agent always moves towards its current local goal until the latter is reached. When it happens next waypoint from the path becomes the local goal and \textsc{ORCA} is re-invoked. It is noteworthy, that when an agent moves to the current goal it may significantly deviate from the original path's segment (due to other moving agents forcing \textsc{ORCA} to do so). Thus, it may appear that the local goal can no longer be reached via the straight-line segment, which is a crucial condition for ORCA to operate appropriately. We detect such patterns and trigger re-planning from the current agent's position to the local goal with \textsc{Theta*}. New waypoints are added to sequence of the local goals as a result. The condition that there is a visible connection between any pair of waypoints now holds and ORCA continues. An illustration of the overall process is presented in~\figurename~\ref{fig:currgoal}.

\subsection{Multi-agent Path Finding for Avoiding the Deadlocks.}

\begin{figure}[t]
    \centering
    \includegraphics[width=0.6\textwidth]{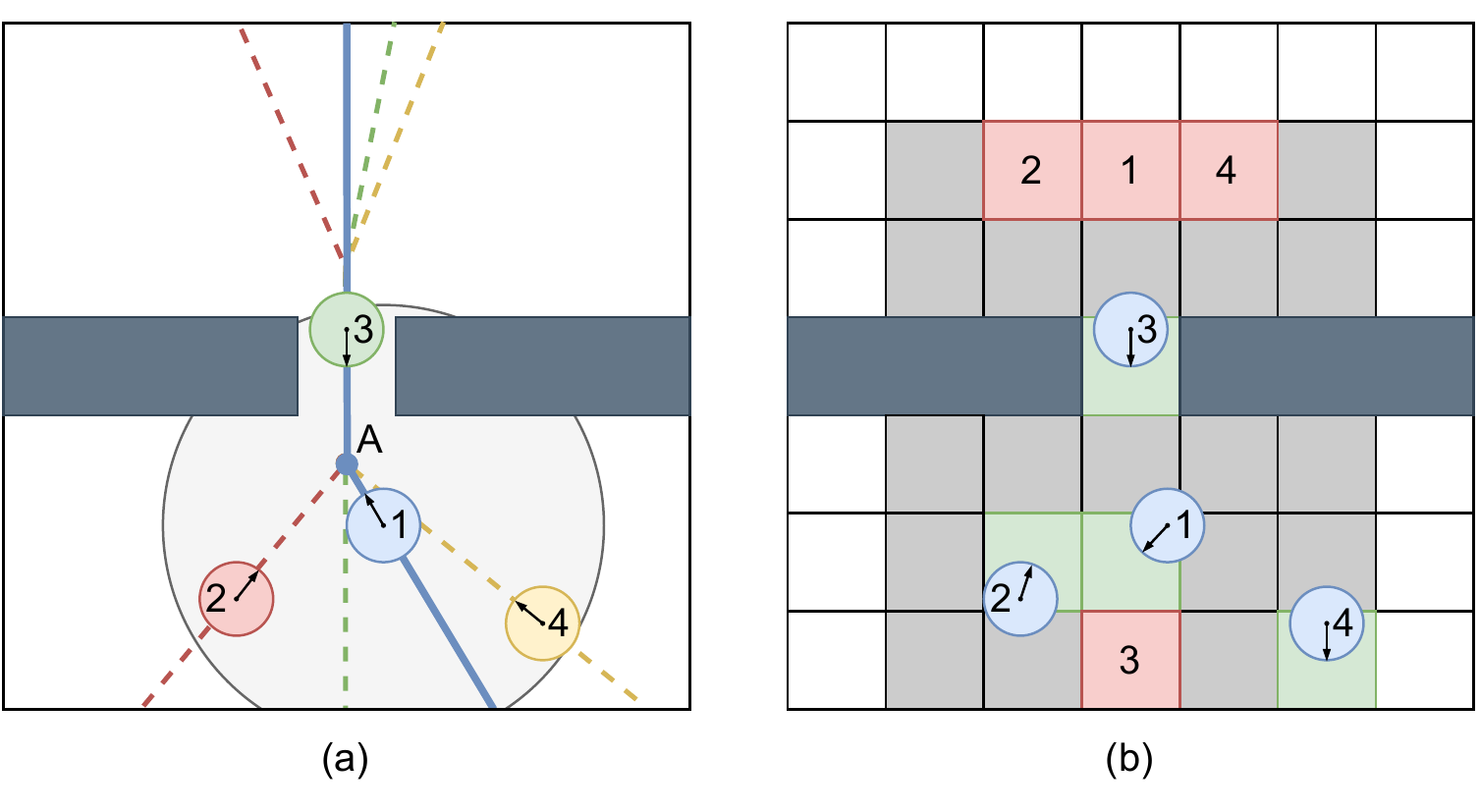}
    \caption{Switching to coordinated mode. (a) Agent 1 detects its local goal, $A$, and 3 other agents within its visibility zone. (b) Based on the current positions and individual paths of agents 1-4 a local grid-map for multi-agent path-finding is identified as well as start and goal locations for each agent.} 
    \label{fig:area}
\end{figure}

Path planning and collision avoidance mechanisms described so far can be attributed as non-cooperative, i.e. each agent pursues its own goal and does not take the goals of other agents into account. In general this can lead to deadlocks. A typical scenario when such a deadlock occurs is depicted on \figurename~\ref{fig:area} on the left. Here all four agents have to come through the tight passage nearly at the same time. In such case \textsc{ORCA} is forced to set the velocity of each agent to zero to avoid the collisions. Thus, no agent is moving and a deadlock occurs. To avoid this we include a multi-agent path finding (MAPF) algorithm into the navigation pipeline with the aim of building a set of locally coordinated collision-free paths. Conceptually, when an agent detects a pattern that leads to a potential deadlock it switches to a coordinated path planning mode and forces its neighboring agents to do so. Upon entering this mode involved agents create a joint conflict-free plan and execute it. After they switch back to the independent execution mode.

Switching to the coordinated mode is triggered if both local goal $A$ and $k$ other agents appear in the visibility range of some agent (here $k$ is the predefined threshold set up by the user). The latter forms a list of coordinated agents that include: itself, its neighboring agents and their neighboring agents as well. All these agents now share the information about their current states (positions and velocities) and individual paths with each other. Thus each agent constructs the same information model and uses it for operation, so no explicit centralized controller is introduced.

Based on the current positions of the agents, the boundaries of the area in which the joint collision-free plan will be built are estimated. This is done as follows. The minimum and maximum $x$- and $y$-positions across all agents are identified. These four coordinates define a square which is inflated by a visibility radius of the agent that has initiated switching to the coordinated mode. This square is translated to the grid which was originally used for the individual path planning. The resulting grid now becomes a local map for multi-agent path finding (it is depicted on~\figurename~\ref{fig:area}b in grey). It might appear that two or more coordinated planning areas are separately constructed by different groups of agents close to each other. In this case those areas are combined into a single one and all agents become members of a single coordinated group. 

After the grid map for multi-agent path finding is constructed each agent chooses its MAPF start and MAPF goal on that grid. The start cell is the one closest to the current location of the agent. If it coincides with the start of another agent or it is blocked by an obstacle a breadth-first search graph traversal is invoked from this cell to find a close un-blocked position which becomes a start. Goal locations are identified in the similar way. First, a current local goal (path's waypoint) is identified. If it coincides with $A$ the next waypoint is chosen. If the selected waypoint lies inside the planning area then the cell which contains it become the MAPF goal. If the waypoint is outside an area, i.e. at least its $x$- or $y$- coordinate is greater/less than the corresponding maximum/minimum values of the planning area, then this coordinate is replaced by the maximum or minimum value of the corresponding coordinate.

After the start and goal positions of the agents are fixed, an appropriate MAPF solver is invoked to obtain a solution, i.e. a set of sequences of moves between the grid cells and, possibly, wait actions, -- one sequence per each agent. For the sake of simplicity we assume that all MAPF actions, i.e. move and wait, are of the uniform duration. This duration is chosen in such a way that the constraints on the agents' maximum moving speed are not violated. In this work we utilize \textsc{Push and Rotate}~\cite{de2014push} for multi-agent path planning algorithm as it is very fast and scales well to large number of agents. We did not choose an optimal solver, such as \textsc{CBS}~\cite{sharon2015conflict}, because of its high computational budget. Using a prioritized planner is also not an option as prioritized MAPF solvers do not guarantee completeness. 

When a MAPF problem is solved each agent starts moving towards its chosen MAPF-start position on a grid (with ORCA collision avoidance activated). When all agents are at their start positions synchronous execution of the plan begins. At this stage no collisions are guaranteed to happen (assuming perfect execution) so ORCA is not used. When all agent reach their MAPF-goals they switch back to the individual mode, i.e. continue path following to their next waypoint on a global path with \textsc{ORCA} as described in previous section.

In case some agent, $i$, which is not involved in the plan execution, gets inside the boundaries of the planning area and interferes with the coordinated agents, i.e. this agent appears in any coordinated agent's visibility range, the execution of the coordinated plan is stopped, the agent $i$ is added to the list of coordinated agents and the plan is recomputed.

Similarly, if an agent $i$, which is involved in the plan execution, gets inside into the visibility zone of an agent $j$, which is also involved in the execution of plan, but belongs to another group of coordinated agents, then these two coordinated groups are merged and re-planning is triggered.

\section{Experimental Evaluation}

\begin{figure}[t]
    \centering
    \begin{subfigure}[b]{0.3\textwidth}
        \includegraphics[width=\textwidth]{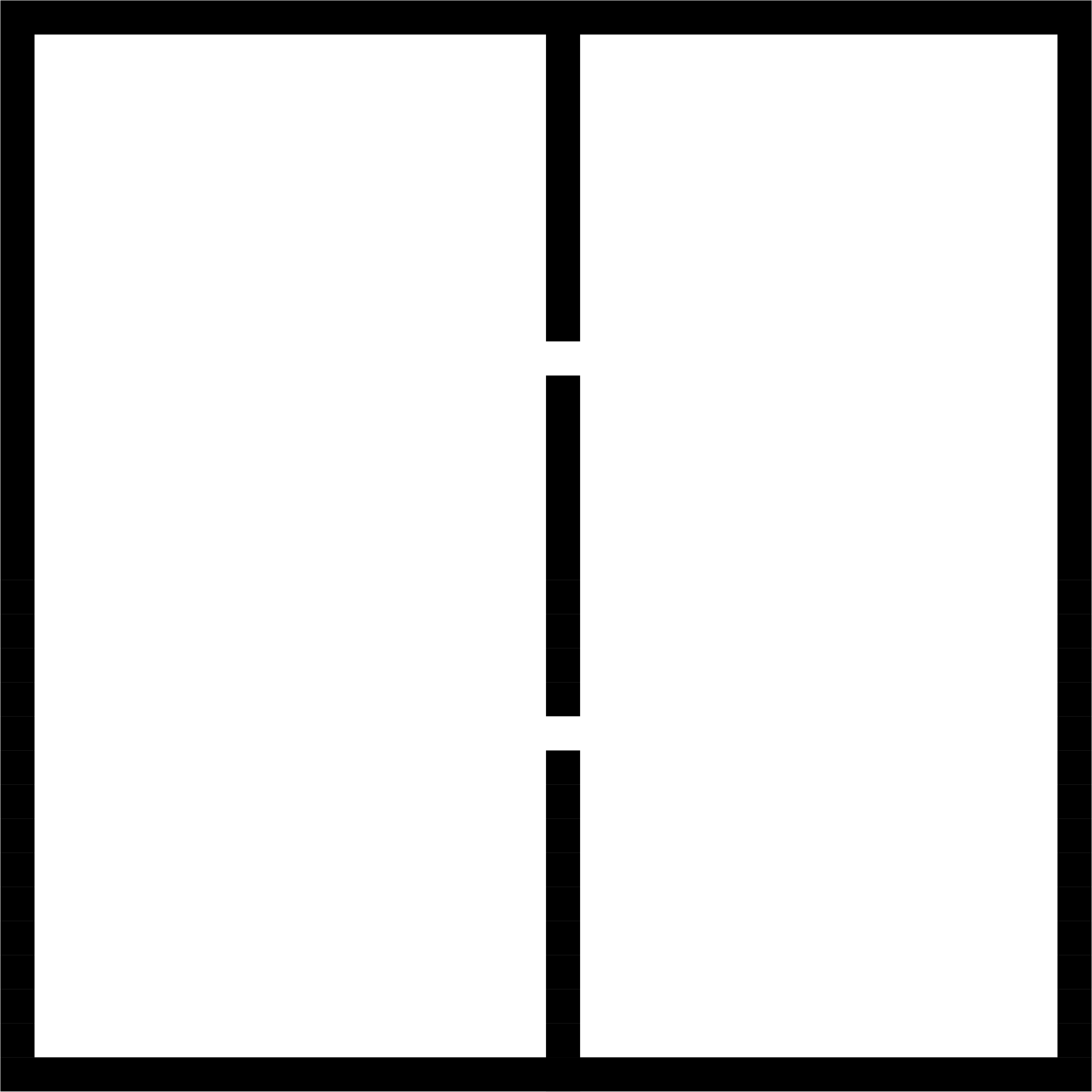}
        \label{fig:map3}
        \caption{}
    \end{subfigure}
    \begin{subfigure}[b]{0.3\textwidth}
        \includegraphics[width=\textwidth]{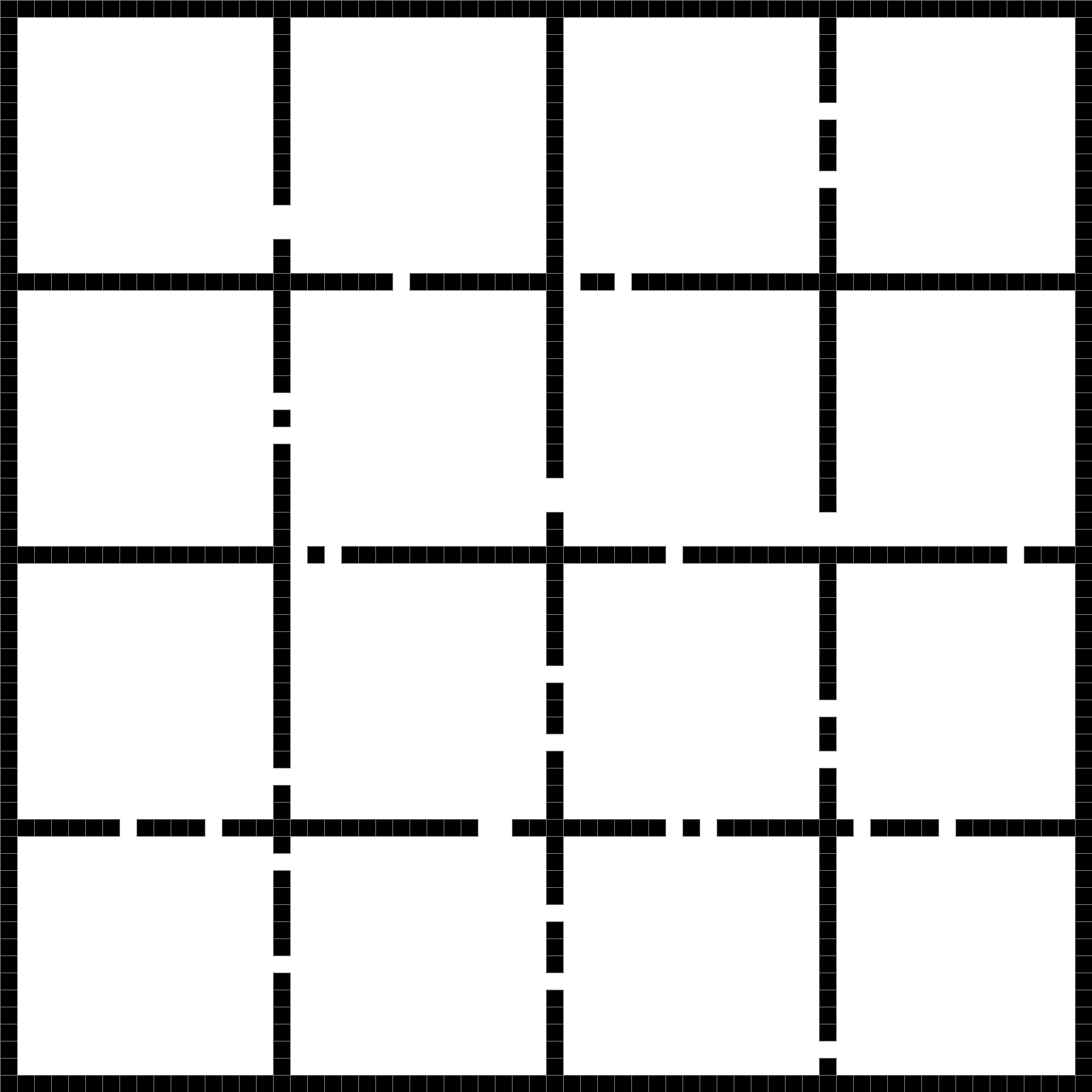}
        \label{fig:map1}
        \caption{}
    \end{subfigure}
    \begin{subfigure}[b]{0.3\textwidth}
        \includegraphics[width=\textwidth]{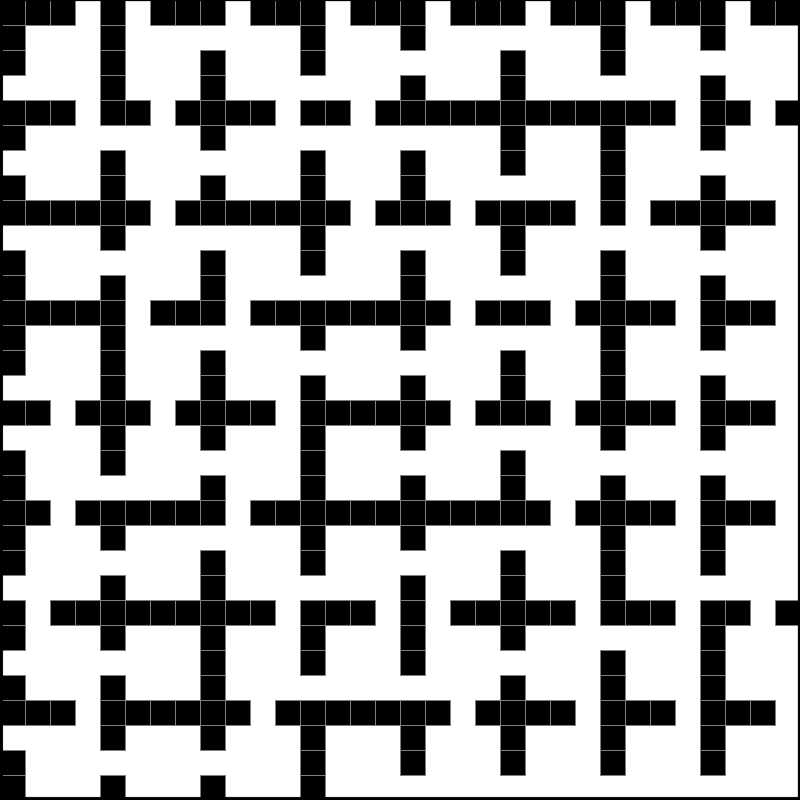}
        \label{fig:map2}
        \caption{}
    \end{subfigure}
    \caption{Maps used for experimental evaluation. (a) Map comprised of two rooms connected by narrow passages. (b) and (c) Maps of the indoor environment composed of numerous rooms connected by narrow passages.}
    \label{fig:maps}
\end{figure}

The proposed method was implemented in C++\footnote{\url{https://github.com/PathPlanning/ORCA-algorithm/tree/Deadlocks}} and its experimental evaluation was carried out on a laptop running macOS 10.14.6 based on Intel Core i5-8259U (2.3 GHz) CPU and with 16 GB of RAM.

We used two types of grid maps, i.e. the maps composed of the free/blocked cells, for the experiments (see~\figurename~\ref{fig:maps}). First, we generated maps that were comprised of two open areas (rooms) separated by a wall with tight passages (doors) in it. The number of passages varied from 1 to 4, thus 4 different maps of that type were generated in total. The overall size of each map was $64 \times 64$. We refer to these maps as to the \texttt{gaps} maps. Second, we took two maps from the \textit{MovingAI}~\cite{stern2019multi} benchmark. They represent indoor environments composed of numerous rooms connected by passages. The first map of that type has a size of $64 \times 64$ and was comprised of 16 rooms (9 of them of the size $15 \times 15$, 1 --  $14\time14$ and the rest -- $14 \times 15$ or $15 \times 14$). The size of the second map was $32 \times 32$ and it was composed of the 64 rooms of the size $3 \times 3$. The latter two maps, referred to as \texttt{rooms} further on, are depicted on~\figurename~\ref{fig:maps}b-c.

\begin{figure}[t]
    \centering
    \begin{subfigure}[b]{0.4\textwidth}
        \includegraphics[width=\textwidth]{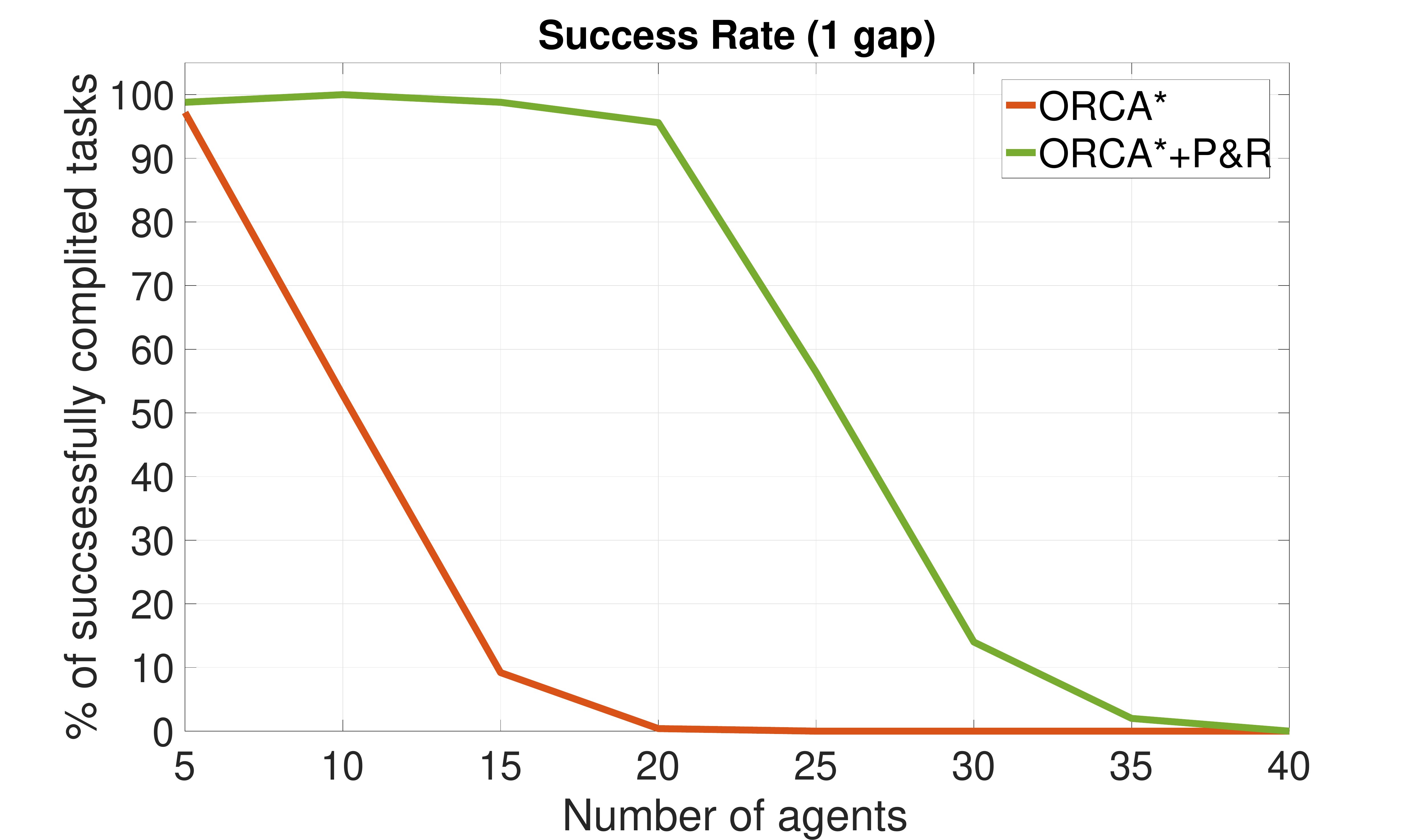}
        \label{fig:SR1}
        % \caption{}
    \end{subfigure}
    \begin{subfigure}[b]{0.4\textwidth}
        \includegraphics[width=\textwidth]{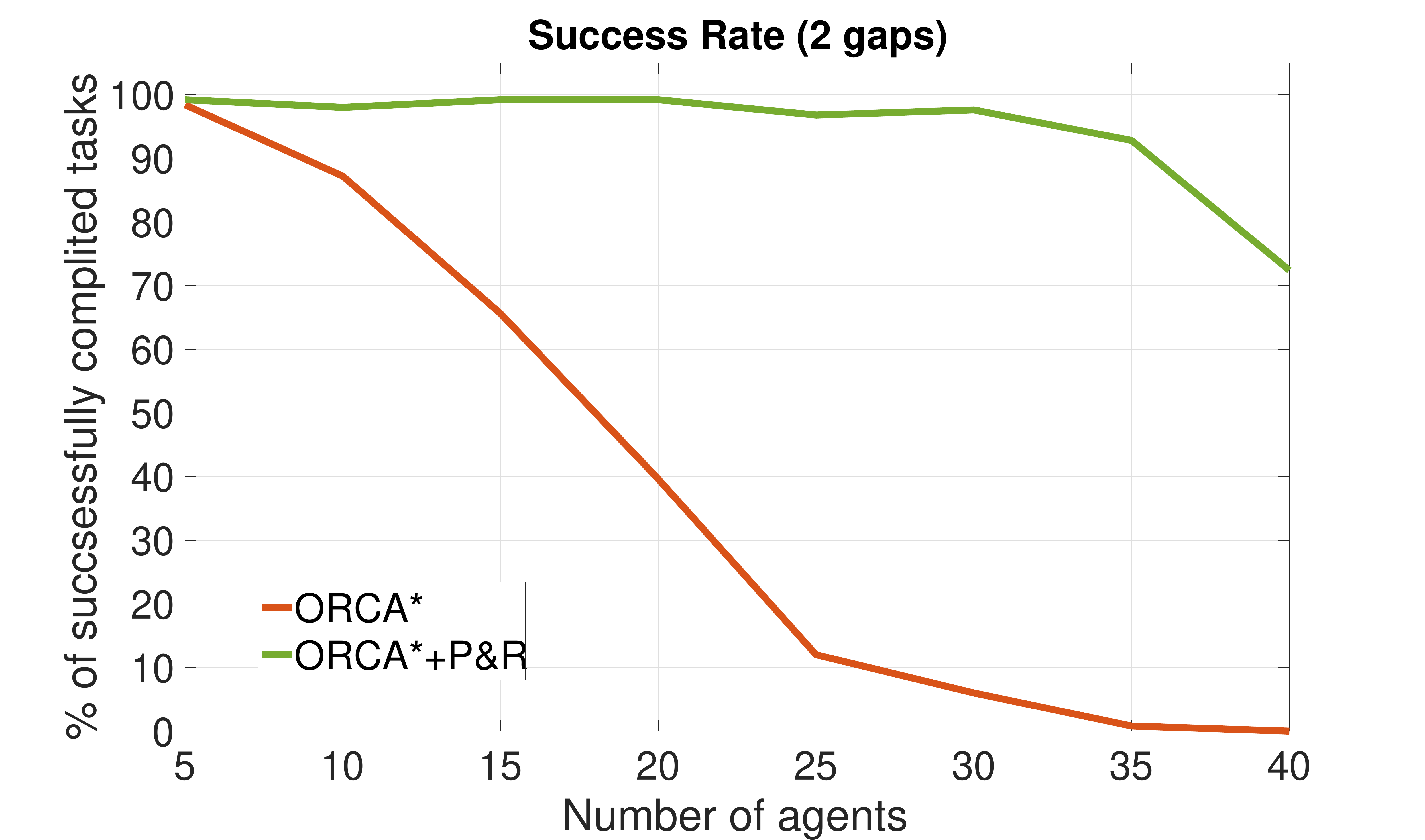}
        \label{fig:SR2}
        % \caption{}
    \end{subfigure}
   
    \begin{subfigure}[b]{0.4\textwidth}
        \includegraphics[width=\textwidth]{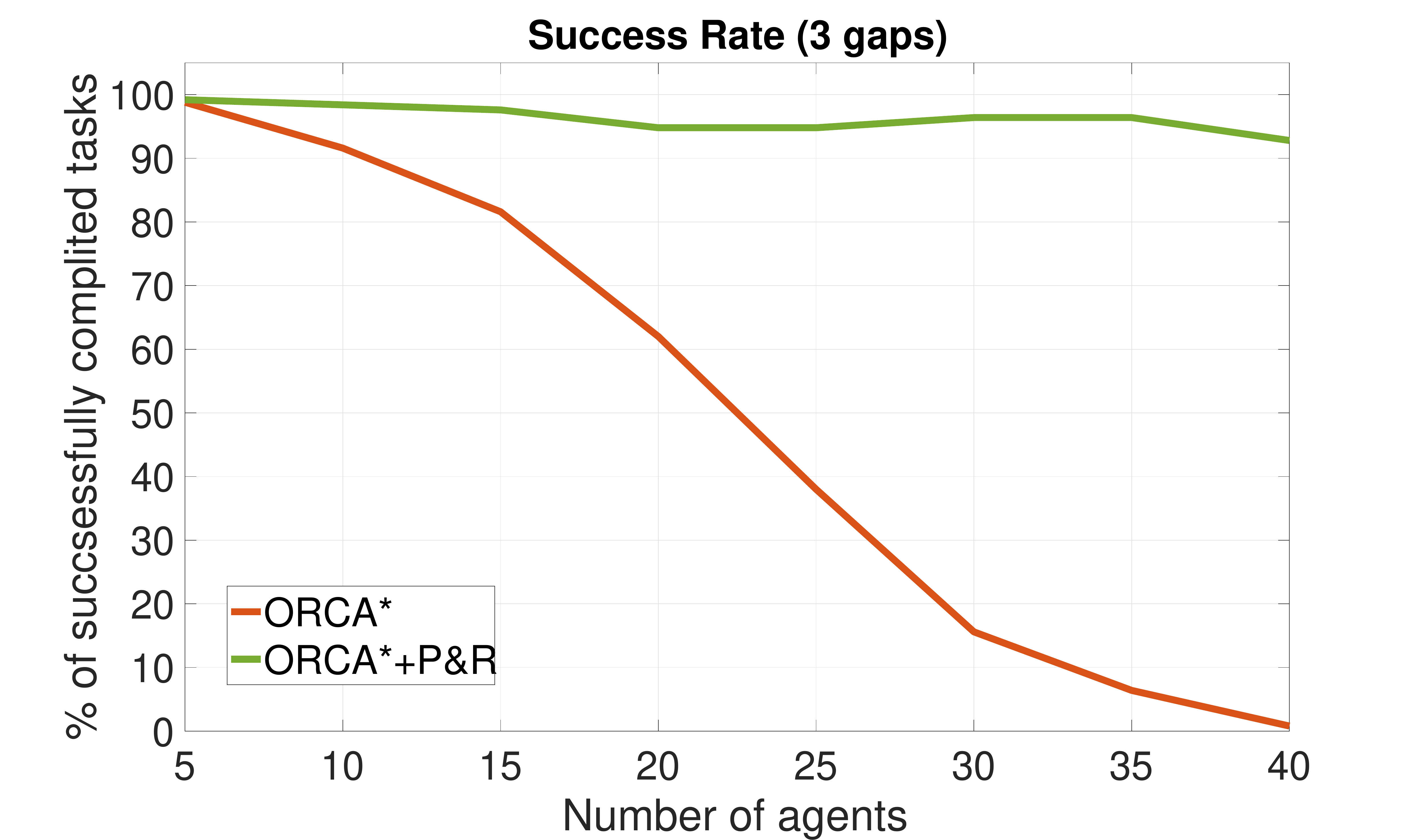}
        \label{fig:SR3}
        % \caption{}
    \end{subfigure}
    \begin{subfigure}[b]{0.4\textwidth}
        \includegraphics[width=\textwidth]{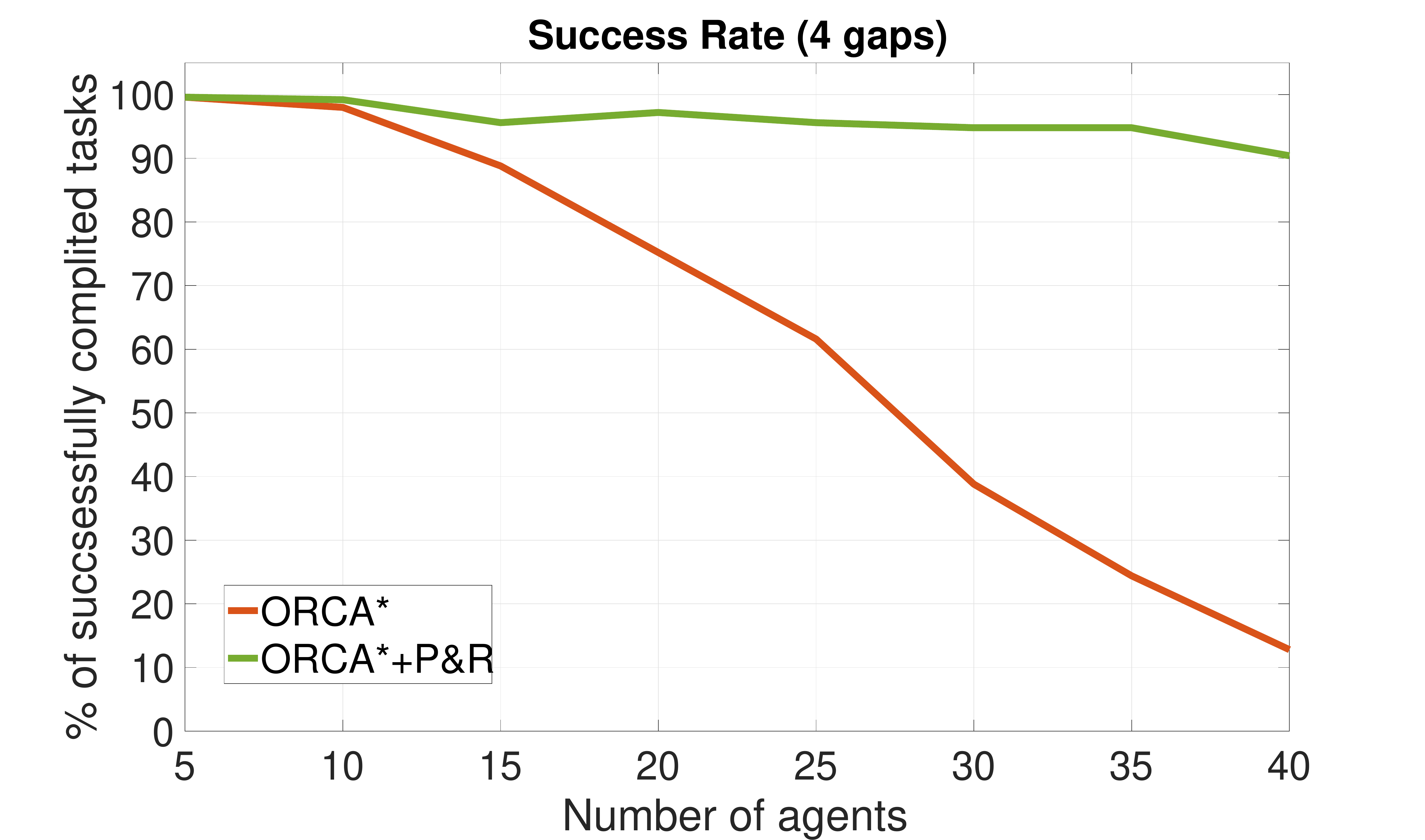}
        \label{fig:SR4}
        % \caption{}
    \end{subfigure}
    
    \caption{Success rate for \texttt{gaps} maps.}
    \label{fig:SRgaps}
\end{figure}

Start/goal locations on each map were generated as follows. For the \texttt{gaps} maps we always generated half of the start/goal locations in the left open area and the other half in the right one. So, to accomplish the mission, agents have to navigate between the two areas via the passages in the wall. For the \texttt{rooms} maps we picked start/goal locations randomly. For each map we created 250 different scenarios. Each scenario is a list of 40 start/goal locations for the agents. While testing the scenarios were used in the following fashion. First, we invoke the algorithm on 5 (first) start/goal pairs of a scenario, then we increased this number by 5 and invoke the algorithm again and so on until the number of agents reaches 40. At this time the scenario was considered to be processed and we moved to the next one.

We compared the suggested approach, referred to as \textsc{ORCA*+P\&R}, to the one that does not utilize deadlock detection and multi-agent path finding, referred to as \textsc{ORCA*}. For both versions we set the radius of each agent to be equal to 0.3 of the cell size. We also introduced an additional 0.19 safe-buffer for individual path finding and collision avoidance. For \textsc{ORCA*+P\&R} the number of neighbours for an agent required to switch to the coordinated mode was set to $k=3$.

At each run the maximum number of simulation steps was limited to 12 800. If by that time the agents fail to reach their goals, i.e. at least one agent was not on its target position or at least one agent collided with another agent or static obstacle, the run was considered to be \textit{failure}. If all agents managed to safely reach their goals the run was acknowledged as \textit{success}. The main metric that we were interested in was the \textit{success rate} i.e. the percentage of tasks that result with  \textit{success} out of all tasks attempted to be solved.

\begin{figure}[t]
    \centering
    \begin{subfigure}{0.4\textwidth}
        \includegraphics[width=\textwidth]{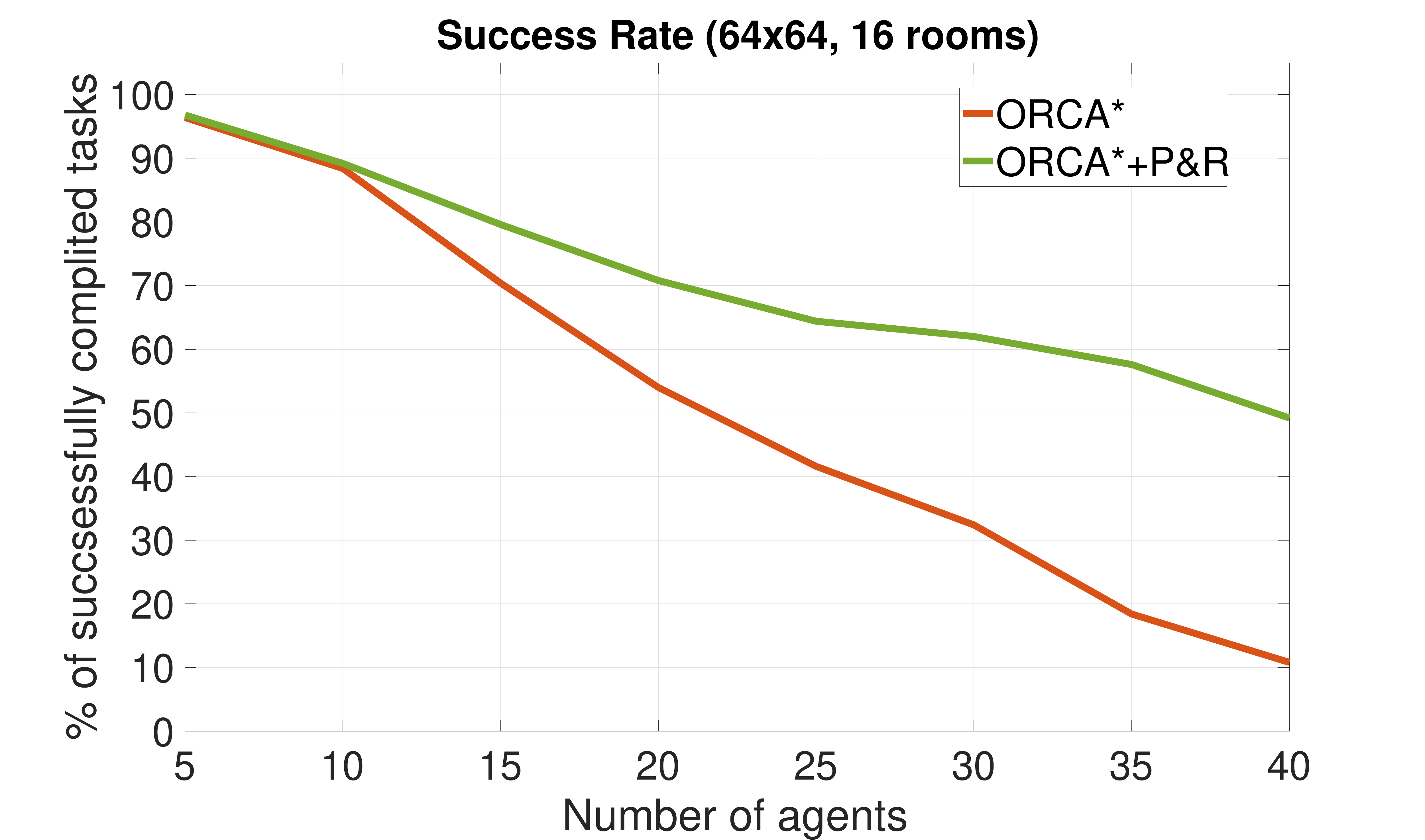}
        \label{fig:room64}
        % \caption{64x64, 16 rooms}
    \end{subfigure}
    \begin{subfigure}{0.4\textwidth}
        \includegraphics[width=\textwidth]{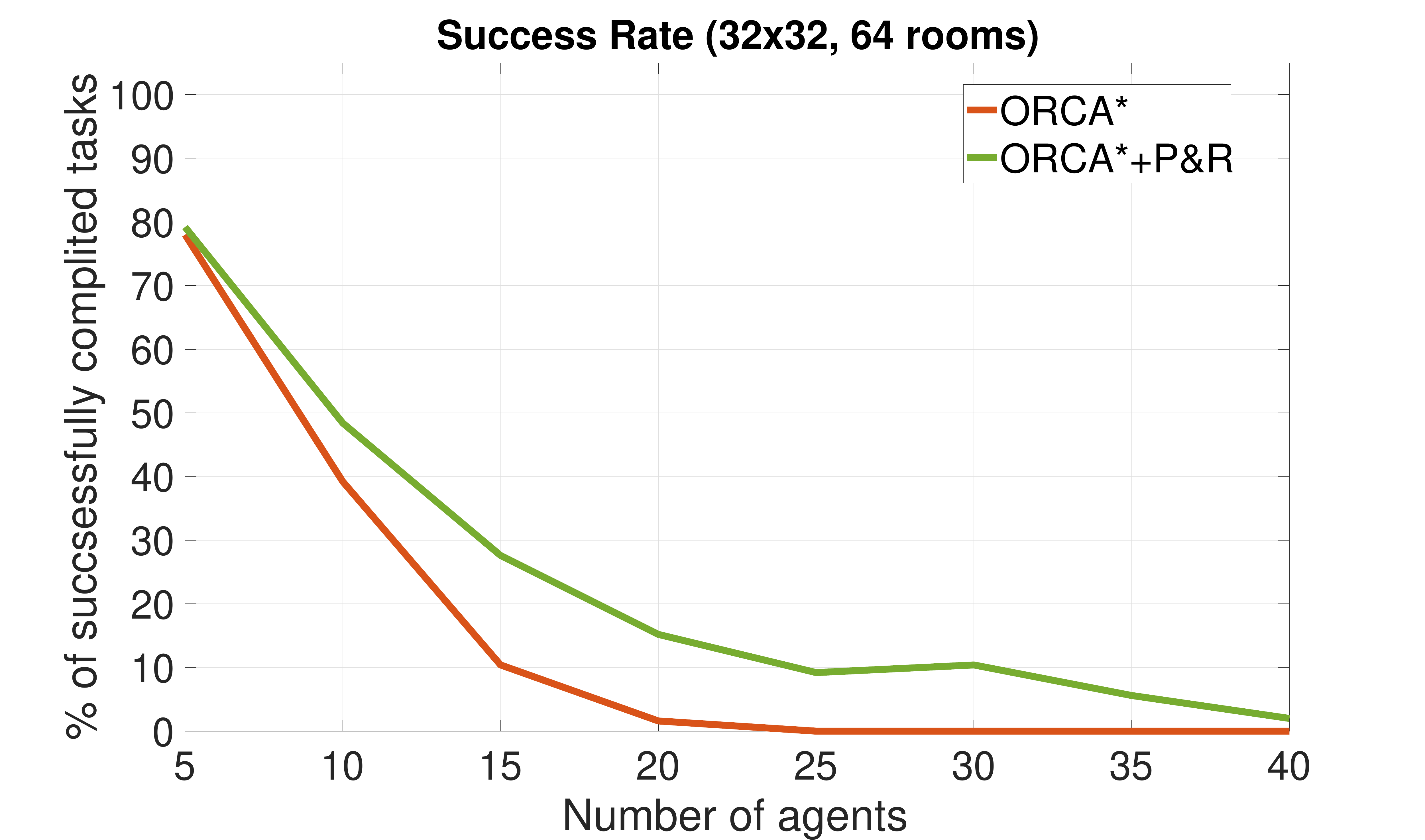}
        \label{fig:room32}
        % \caption{32x32, 64 rooms}
    \end{subfigure}
    
    \caption{Success rate for \texttt{rooms} maps.}
    \label{fig:SRrooms}
\end{figure}

Success rates for \textsc{ORCA*+P\&R} and \textsc{ORCA*} are shown on~\figurename~\ref{fig:SRgaps} and \ref{fig:SRrooms}. As one can see the selected environments are very challenging for \textsc{ORCA*} and its success rate is very low for large number of agents on all the maps. The most challenging environments are the one with  1 passage on a \texttt{gaps} map and the one with 64 rooms on \texttt{rooms} map. As expected on all maps \textsc{ORCA*+P\&R} successfully solved profoundly more instances compared to the baseline. For example, for the challenging \texttt{gaps} map with 1 passage the success rate for \textsc{ORCA*+P\&R} was 95\% for 20 agents while the one for \textsc{ORCA*} was 0\%. The difference in success rate for \texttt{rooms} maps is also clearly visible but less pronounced. We hypothesise that the main reason for that, as well as the reason for \textsc{ORCA*+P\&R} not to achieve 100\% success in general, is the imposed time limit. The plans created by the \textsc{Push and Rotate} algorithm may have contained numerous actions and that prevented the agents to reach their goals before the timestep limit exhausted. One of the possible solutions to avoid this will be substituting  \textsc{Push and Rotate} to another MAPF solver which creates plans containing less actions. However the solvers that are explicitly aimed at minimizing duration of the plan have much higher computational budget, so finding a right substitution is a challenging task for future research.

Overall, the results of the experiments clearly show that adding a locally-confined multi-agent path finding into the navigation pipeline significantly increase the chance that all agents will reach their goal locations and mission, thus, be accomplished.

\section{Conclusion}

In this paper we have studied a multi-agent navigation problem and suggested a decentralized approach that supplements the individual path planning and collision avoidance with the deadlock detection and multi-agent path finding. We implemented the proposed navigation pipeline and compared it the to the baseline, showing that adding the aformentioned components significantly increases the chances of agents safely arriving to the target destinations even in the congested environments with tight passages. An appealing direction of future work is the advancement of the deadlock detection procedure by making it less ad-hoc, as well as experimenting with different multi-agent path finding algorithms. Another direction for future research is evaluating the suggested approach on real robots.

\subsubsection{Acknowledgements}
The reported study was funded by RFBR and BRFBR, project number 20-57-00011. We would also like to thank Ilya Ivanashev for his implementation of \textsc{Push and Rotate}.

\bibliographystyle{splncs04}
\bibliography{bibl}

\end{document}